\documentclass[twocol]{ametsocV5_navid}

\usepackage{amsmath,amsfonts,amssymb,bm}
\usepackage{mathptmx}
\usepackage{newtxtext}
\usepackage{newtxmath}
\usepackage{comment}
\usepackage{subcaption}
\usepackage[normalem]{ulem}
\usepackage{printlen}
\usepackage{doi}

\PassOptionsToPackage{hyphens}{url}\usepackage{hyperref}
\hypersetup{
	breaklinks,
	colorlinks=true,
	linkcolor=blue,
  urlcolor=Purple,
	citecolor=Purple,
 }

\graphicspath{{figs/}}

\newcommand{\lon}{\mathrm{lon}}
\newcommand{\lat}{\mathrm{lat}}
\newcommand{\fmin}{f_{\min}}
\newcommand{\fmax}{f_{\max}}
\newcommand{\var}{\mathrm{var}}

\title{Intrinsic oceanic decadal variability of upper-ocean heat content}

\authors{Navid C. Constantinou\correspondingauthor{Navid Constantinou,\\ navid.constantinou@anu.edu.au} and Andrew McC. Hogg}

\affiliation{Research School of Earth Sciences and ARC Centre of Excellence for Climate Extremes,\\ Australian National University, Canberra, Australian Capital Territory, Australia}

\abstract{Atmosphere and ocean are coupled via air--sea interactions. 
The atmospheric conditions fuel the ocean circulation and its variability, but the extent to which ocean processes can affect  the atmosphere at decadal time scales remains unclear. 
In particular, such low-frequency variability is difficult to extract from the short observational record, meaning that climate models are the primary tools deployed to resolve this question. 
Here, we assess how the ocean's intrinsic variability leads to patterns of upper-ocean heat content that vary at decadal time scales. 
These patterns have the potential to feed back on the atmosphere and thereby affect climate modes of variability, such as El Ni\~no or the Interdecadal Pacific Oscillation. 
We use the output from a global ocean--sea ice circulation model at three different horizontal resolutions, each driven by the same atmospheric reanalysis. 
To disentangle the variability of the ocean's direct response to atmospheric forcing from the variability due to intrinsic ocean dynamics, we compare model runs driven with inter-annually varying forcing (1958-2018) and model runs driven with repeat-year forcing. 
Models with coarse resolution that rely on eddy parameterizations, show \emph{(i)}~significantly reduced variance of the upper-ocean heat content at decadal time scales and \emph{(ii)}~differences in the spatial patterns of low-frequency variability compared with higher resolution models. 
Climate projections are typically done with general circulation models with coarse-resolution ocean components. 
Therefore, these biases affect our ability to predict decadal climate modes of variability and, in turn, hinder climate projections. 
Our results suggest that for improving climate projections, the community should move towards coupled climate models with higher oceanic resolution.}

\begin{document}

\maketitle

%
%
%

%

\section{Introduction}

The atmosphere and ocean communicate at the sea surface via air--sea interactions allowing the exchange of heat, freshwater and momentum. 
There is clear evidence that the atmosphere affects the ocean, since it is responsible for setting up the wind stress and the meridional buoyancy gradients that fuel the ocean's mean circulation and its variability. 
At high frequencies ($\sim$weekly), the atmosphere's synoptic variability appears at large length scales ($\sim$1000~km). 
The ocean's large heat capacity acts as memory since it results in suppressing or ``integrating out'' the high-frequency atmospheric variability, enhancing power in longer timescales ($\sim$decadal) of the coupled atmosphere--ocean climate system. 
This is the ``null hypothesis'' for explaining the observed red spectrum of the climate variability. 
According to the null hypothesis, the ocean's low-frequency response results solely from the atmosphere's high-frequency synoptic variability (which, from the ocean's viewpoint, appears to leading order as stochastic excitation), and the role of the ocean is to introduce the slower time scale (due to its larger heat capacity) so that the atmosphere's stochastic nudging becomes a red-noise process inducing power at low frequencies \citep{Hasselmann-1976, Frankignoul-Hasselmann-1977}. 
The null-hypothesis paradigm forms the basis of many air--sea coupling studies up to date \citep[e.g.,][]{Barsugli-Battisti-1998}.

Recent work revealed that the null-hypothesis paradigm is only one part of the story. 
It has been shown that the variability of air--sea heat fluxes over ocean regions outside the tropics with high mesoscale eddy activity shows a different character from the null-hypothesis prediction: air--sea heat fluxes are not predominantly controlled by the atmosphere in those regions, but rather the other way around \citep{Bryan-etal-2010, Buckley-etal-2014, Buckley-etal-2015, Bishop-etal-2017, Small-etal-2019, Small-etal-2020}. 
The question remains whether the ocean's \emph{intrinsic variability} can enhance power at inter-annual frequencies, thus allowing for the potential to affect the climate at decadal time scales.\footnote{By ``intrinsic ocean variability'' here we mean the variability that rises due to oceanic dynamical processes and not as a direct response of the ocean to the atmospheric forcing; we will  refer to the latter as ``forced ocean variability.''} Is it possible that the ocean's variability can enhance the low-frequency variability of air--sea heat fluxes through the ocean's capacity to create sea-surface temperature variability at decadal time scales? Furthermore, to what extent does such low-frequency variability can affect the atmosphere \citep{Vallis-2010}? We attempt here to tackle both of these questions. 
Thus, our main objective in this paper is to assess whether and how the ocean's intrinsic variability affects the climate variability on decadal time scales.

What could be the cause of such ocean-generated decadal variability? One hypothesis is that the high-frequency atmospheric variability acts as noise thus enabling the ocean to undergo transitions between different close-to-equilibrium states. 
Take, for example, a particle in a double-well potential. 
Without any external noise the particle sits at the the bottom of either of the potential wells. 
With some noise, the particle starts wandering around the bottom of the well and, when the noise becomes sufficiently strong, the particle may jump between wells and end up at the bottom of the neighboring potential-well. 
Such transitions from one well to another occur less frequently than oscillations due to noise and thus appear as low-frequency variability of the particle's position. 
Turbulent flows often support large-scale coherent features; such features are ubiquitous in the ocean, e.g., the Gulf Stream, Kuroshio Current, the Antarctic Circumpolar Current. 
The different states of these coherent features manifest as equilibria of the dynamics that the flow-statistics obey (e.g., \citet{Farrell-Ioannou-2008-baroclinic, Parker-Krommes-2013, Constantinou-etal-2014, Constantinou-2015-phd, Farrell-Ioannou-2019-book}). 
In general, flows tend to ``wander around'' such equilibria (similarly to how the  particle wanders around the bottom of the well when there is noise). 
When the turbulent flow attractor possesses more than one stable equilibria (``bi-stability'' -- double-well), then transitions between the such equilibria can occur \citep{Kimoto-Ghil-1993a, Qiu-Miao-2000, Pierini-2006, Pierini-etal-2009, Parker-Krommes-2013, Constantinou-2015-phd}. 
These transitions typically occur at much longer time scales than the time scale the flow exhibits and, thus, appear as lower-frequency variability.

In the atmosphere, low-frequency variability of precisely this character was demonstrated with the seminal experiments by \citet{James-James-1992}. 
Those experiments revealed a so-called ``ultra-low-frequency variability,'' which was shown to come about from transitions between a two-jet state, with a subtropical jet distinct from an eddy-driven mid-latitude jet, and a single or merged jet state. 
Therefore, in \citet{James-James-1992}'s experiments, the intrinsic nonlinearity of the atmosphere became the source of low-frequency variability. 
Notably, \citet{James-1998} argued that models with parameterizations cannot capture the low-frequency variability that was seen in the experiments of \citet{James-James-1992}. 
Recent studies demonstrated the occurrence of intrinsic low-frequency variability due to bi-stability of the turbulent attractor in barotropic models \citep{Bouchet-etal-2019-jettransisions, Simonet-etal-2021}.

In the ocean, the primary patterns of variability occur on smaller scales ($\sim$100~km) and lower frequencies ($\sim$months) than the atmosphere, through the creation of mesoscale eddies predominantly via baroclinic instability. 
However, the influence of nonlinear processes in creating \emph{intrinsic} variability at lower frequencies (defined in this paper to be approximately decadal) is difficult to evaluate either from observations or coupled climate models, because variability of the ocean circulation is also \emph{forced} from the atmosphere. 
The superposition of forced and intrinsic variability requires a targeted approach to the problem. 

\citet{Hogg-etal-2005} found that idealized eddy-rich ocean models show high intrinsic variability, which is greatest in eddying regions and can influence the atmospheric variability \citep[see also][]{Hogg-etal-2006, Martin2020, Martin-etal-2021}. 
Bi-stability in the oceanographic context has been discussed by \citet{Deshayes-etal-2013} and \citet{Aoki-etal-2020}. 
Other mechanistic and dynamical explanations have been proposed as candidates for inducing low-frequency variability in the ocean: the turbulent oscillator \citep{Berloff-etal-2007}, which involves eddy--mean flow feedback between meridional eddy fluxes and the strength of a quasi-zonal jet. 
These proposed mechanisms depend largely on resolving nonlinear scales of motion, principally ocean eddies. 
Climate projections, on the other hand, are routinely done with general circulation models that use a ``laminar ocean,'' i.e., ocean components whose lateral resolution is too coarse to resolve ocean eddies and instead rely on eddy-parameterizations \citep{Hewitt-etal-2020}. 
What remains outstanding and forms the basic question addressed in this paper is whether the ocean's intrinsic mesoscale flow has the potential to affect the large-scale patterns of sea-surface temperature variability at decadal time scales.

A series of studies aimed to better understand the forced and intrinsic response of the ocean using numerical experiments utilized a large ensemble of global ocean--sea ice model simulations (OCCIPUT project; \href{http://meom-group.github.io/projects/occiput}{http://meom-group.github.io/projects/occiput}). 
These simulations were all initialized with slightly different initial conditions and they were run at a resolution that was able to partially resolve some of the ocean eddies \citep{Penduff-etal-2011}. 
The chaotic nature of the ocean's intrinsic variability in certain regions leads to a spread in the solutions, while in other regions where the forced component dominates, the oceanic response among the ensemble members revealed more similarities. 
These results suggested that intrinsic low-frequency variability is significant in eddy-present models and can be as large as 80\% of all observed variability in regions of high ocean-eddy activity \citep[see also][]{Penduff-etal-2018, Leroux2018}. 
\citet{Serazin-etal-2017} used the OCCIPUT ensemble dataset to quantify the forced and intrinsic footprint of the ocean heat content at decadal time scales. 
They found that the ocean-heat-content variance at decadal time scales increased in regions of high eddy activity. 
Here, we will focus on the upper-ocean heat content and study how low-frequency variance changes as we go from models that parameterize ocean eddies towards models that resolve the eddies. 
We furthermore determine the main patterns of low-frequency intrinsic variability that ocean dynamics induce on the upper-ocean heat content using an empirical orthogonal function-analysis and study how these patterns change with increasing model resolution.

One approach to decompose forced from intrinsic ocean variability is to use long, eddy-rich global ocean--sea ice model runs  \citep{Serazin-etal-2015}. 
To distinguish the forced from the intrinsic component, we can compare ocean--sea ice models driven by realistic, inter-annually varying atmospheric forcing with models driven by a modified atmospheric forcing that does not vary at sub-annual frequencies \citep{Stewart-etal-2020}. 
Given that in the latter experiments the forcing of the models is repeated every year, and thus does not vary at time scales longer than one year, any ocean variability we observe at decadal time scales is attributed to intrinsic variability. 
In this paper, we will undertake this approach to disentangle the forced from the intrinsic ocean's response, and also to compare how the intrinsic variability of the upper ocean heat content varies as we refine our model's lateral resolution to better resolve ocean eddies.

In what follows, we describe in detail the datasets we use and the methods we apply to understand the low-frequency variability of upper-ocean heat content (section~\ref{sec:methods}). 
Our results are then presented in section~\ref{sec:results}, which is partitioned in two parts: section~\ref{sec:results}\ref{sec:varmaps} studies the intrinsic versus forced low-frequency variance content across models with different lateral resolutions and section~\ref{sec:results}\ref{sec:eofs} presents global and regional empirical-orthogonal functions analysis to understand the main modes of spatial and temporal low-frequency variability. 
We conclude with a discussion in section~\ref{sec:discussion}.

\section{Methods}
\label{sec:methods}

We use the ocean-sea ice model ACCESS-OM2 at three different horizontal resolutions \citep{Kiss-etal-2020}: eddy-rich at 0.10$^\circ$, eddy-present at 0.25$^\circ$, and at 1$^\circ$ with parameterized eddies.\footnote{Note, that `eddy-present'/`eddy-rich' characterizations do not derive from any strict definition. 
We use them here implying that an eddy-rich model resolves eddies at most latitudes, while an eddy-present model only resolves eddies in the tropics \citep{Hallberg-2013, Hewitt-etal-2020}.} The model is forced with 3-hourly output from the atmospheric reanalysis JRA55-do dataset \citep{Tsujino-etal-2018}. 
To separate the intrinsic from the forced component of the oceanic flow, we use two different forcing strategies: \emph{(i)}~3-hourly inter-annually-varying forcing (IAF) from 1958-2018 which is repeated to yield multiple $\sim60$-year cycles, and \emph{(ii)}~3-hourly repeat-year forcing (RYF), defined to be time-varying forcing from a single year (May 1990--April 1991; in which major climate mode indices are neutral) that keeps repeating itself \citep{Stewart-etal-2020}. 
Note that the IAF cycles are combined together in one long time-series for analysis. 
None of the RYF forcing fields used to drive the ocean have power at time scales longer than one year. 
Thus, in RYF experiments, any variability at time scales longer than one year is attributed to intrinsic oceanic processes. 
More details about the datasets used here, including the length of the model runs, are reported in table~\ref{table1}.

\begin{table*}[t]
\caption{Record length and time period of the satellite altimetry observational dataset by the Copernicus Marine and Environment Monitoring Service (CMEMS) and model output used in this study.
Simulations are indicated by their resolution (1$^\circ$; 0.25$^\circ$ or 0.1$^\circ$) and forcing method (IAF $=$ Interannual Forcing; RYF $=$ Repeat Year Forcing).}\label{table1}
\begin{center}
\begin{tabular}{ccccc}
\hline\hline
abbreviation &  & period of & record length & frequency\\
 &  &  forcing  &  (years) & \\
 \hline
CMEMS & (satellite) & Jan 1993 - Dec 2019 & 27	 & monthly \\
\\
IAF 1$^\circ$ & JRA55-do v1.3 & Jan 1958 - Dec 2017 & 240 (4 cycles) & yearly UOHC / monthly ssh  \\
IAF 0.25$^\circ$ & JRA55-do v1.3 & Jan 1958 - Dec 2017 & 258 (>4 cycles) & yearly UOHC / monthly ssh \\
IAF 0.10$^\circ$ & JRA55-do v1.4 & Jan 1958 - Dec 2018 & 183 (3 cycles) & monthly \\
RYF 1$^\circ$ & JRA55-do v1.3 & May 1990 - Apr 1991 & 260 & monthly \\
RYF 0.25$^\circ$ & JRA55-do v1.3 & May 1990 - Apr 1991 & 250 & monthly \\
RYF 0.10$^\circ$ & JRA55-do v1.3 & May 1990 - Apr 1991 & 220 & monthly \\
\hline
\end{tabular}
\end{center}
\end{table*}

The atmosphere feels the ocean primarily as a thermal boundary condition and, thus, one may argue that to quantify the effect of how the ocean feeds back on the atmosphere, we should look at sea-surface temperature. 
However, in ocean--sea ice models, such as that used here, the prescribed atmosphere does not respond to oceanic heat fluxes. 
Therefore, the atmosphere has effectively infinite heat capacity and the  sea-surface temperature is, to a large degree, ``slaved'' to the imposed atmospheric state \citep{Hyder2018}. 
The nature of this forcing complicates how we can disentangle the forced from the intrinsic oceanic response. 
To overcome this impasse, we will use two other flow quantities to quantify ocean's feedback back to the atmosphere and the global climate at decadal time scales: the sea-surface height (SSH) and the upper-ocean heat content, e.g., conservative temperature integrated over, e.g., the top 50 meters of the ocean. 
By integrating over the top 50 meters of the ocean, we alleviate the `slaving' effect of the sea-surface temperature to the forcing fields that drive the model.

The sea-surface height is intimately related to (albeit not a direct measure of) upper-ocean heat content. 
But more importantly, the observational satellite altimetry record since 1993 allows an estimate of the low-frequency variability from observations, which can be used to ground-truth the models. 
Of course, the observational altimetry record is short (27 years) and, therefore, one should be cautious while interpreting estimates of variability at decadal time scales or longer from such a short record. 
However, comparison with observations enables us to evaluate how well the model captures variability at time scales shorter than 27 years.

In summary, we use output from models with 1$^\circ$, 0.25$^\circ$, and 0.10$^\circ$ horizontal resolution and also the sea-surface height observations from the gridded altimetry CMEMS dataset at 0.25$^\circ$ resolution. 
We first interpolate \emph{all} flow fields to a regular longitude and latitude grid horizontal resolution of 1$^\circ$. 
The motivation behind this choice is that we want to coarsen the resolution to a common grid, so that the baseline resolution of the data processing is equivalent.

The upper-ocean heat content, $\mathcal{H}$, is taken, here, as the heat content over the top 50 meters of the ocean,
\begin{equation}
\mathcal{H}(\lon, \lat, t) = \rho_0 c_p \int_{-50\,\textrm{m}}^\eta T(\lon, \lat, z, t)\, \mathrm{d}z \, ,
\end{equation}
where $T$ is the conservative temperature from the model, $\eta$ is the sea surface height, $\rho_0 = 1035.0\;\textrm{kg}\,\textrm{m}^{-3}$ is the mean density of seawater, and $c_p = 3992.1\;\textrm{J}\,^\circ\textrm{K}^{-1}\,\textrm{kg}^{-1}$ the specific heat capacity of seawater. 
The choice of defining upper-ocean heat content over the first 50$\;$m of the ocean may seem arbitrary; a more natural definition would be to define the upper-ocean heat content over the mixed-layer depth. 
However, such a mixed-layer-depth-dependent definition would create difficulties for comparing across different models with differing lateral resolution and mixed-layer biases. 
We therefore choose a fixed depth for defining upper-ocean heat content and keep in mind that this may result in an underestimate of upper-ocean heat content for regions where the mixed layer is deeper, and an overestimate of upper-ocean heat content in regions where the mixed layer is shallower.

After coarsening all flow fields, we proceed with a frequency decomposition. 
For example, the upper-ocean heat content is decomposed in frequencies via a Fourier transform as,
\begin{equation}
 \hat{\mathcal{H}}(\lon, \lat, f) = \int \mathcal{H}(\lon, \lat, t)\,\mathrm{e}^{-2\pi\mathrm{i} f t}\, \mathrm{d}t \, .
 \label{eq:Hhat}
\end{equation}

We denote the low-frequency component to be anything that corresponds to frequencies smaller than $\fmax = (1.5\,\textrm{years})^{-1}$ (to ensure we exclude the dominant peak of the seasonal cycle) and also larger than a low-frequency cut-off, $\fmin$, i.e.,
\begin{equation}
\hat{\mathcal{H}}_{\mathrm{LF}} = 
     \begin{cases}
       \hat{\mathcal{H}} & \quad \text{if } \fmin < f < \fmax,\\
       0 &\quad\text{otherwise.} \\ 
     \end{cases}
     \label{eq:Hhat_LF}
\end{equation}
The lower frequency cut-off, $\fmin$, in equation~\eqref{eq:Hhat_LF} is chosen to be small enough to include decadal time scales, while excluding any long-term model drift. 
Given that we also want to compare with the satellite altimetry record (which is only 27-years long), we choose $\fmin = (25\,\textrm{years})^{-1}$.

We can reconstruct the low-frequency component of our signal via an inverse Fourier transform of the frequency-decomposition in equation~\eqref{eq:Hhat_LF}, 
\begin{equation}
 \mathcal{H}_{\mathrm{LF}} = \int \hat{\mathcal{H}}_{\mathrm{LF}}\,\mathrm{e}^{2\pi\mathrm{i} f t}\, \mathrm{d}f \, .
 \label{eq:H_LF}
\end{equation}
In view of Parseval's theorem, the low-frequency variance is then given by:
\begin{align}
\var{[\mathcal{H}_{\rm LF}]}(\lon, \lat) & = \frac1{\Delta t} \int \mathcal{H}_{\mathrm{LF}}^2\, \mathrm{d}t = \frac1{\Delta t} \int_{\fmin}^{\fmax} |\hat{\mathcal{H}}_{\mathrm{LF}}|^2\, \mathrm{d}f \, ,
\label{eq:varHLF}
\end{align}
where $\Delta t$ is the duration of the time-series of the signal. 
Similar frequency decomposition and quantification of the low-frequency variance as in equations~\eqref{eq:Hhat}, \eqref{eq:Hhat_LF}, and~\eqref{eq:varHLF} is also performed for the sea-surface height as obtained both from model output and from observations, yielding an estimate of $\var{[{\rm SSH}_{\rm LF}]}$. 
To identify the spatial and temporal patterns of variability, we perform empirical orthogonal function-analysis on the low-frequency component of upper-ocean heat content~$\mathcal{H}_{\rm LF}$.

\section{Results}
\label{sec:results}

\subsection{Low frequency variability}
\label{sec:varmaps}

We first look into the low-frequency variability of sea-surface height and compare it across models with inter-annual or repeat-year forcing, and across the three horizontal resolutions. 
As discussed above, although sea-surface height is not a direct measure of the thermal forcing that the ocean feeds back to the atmosphere, the observed satellite altimetry record can provide a ground-truth for the inter-annually forced model.

\begin{figure*}[t]
 \noindent\centering\includegraphics[width=\textwidth, angle=0]{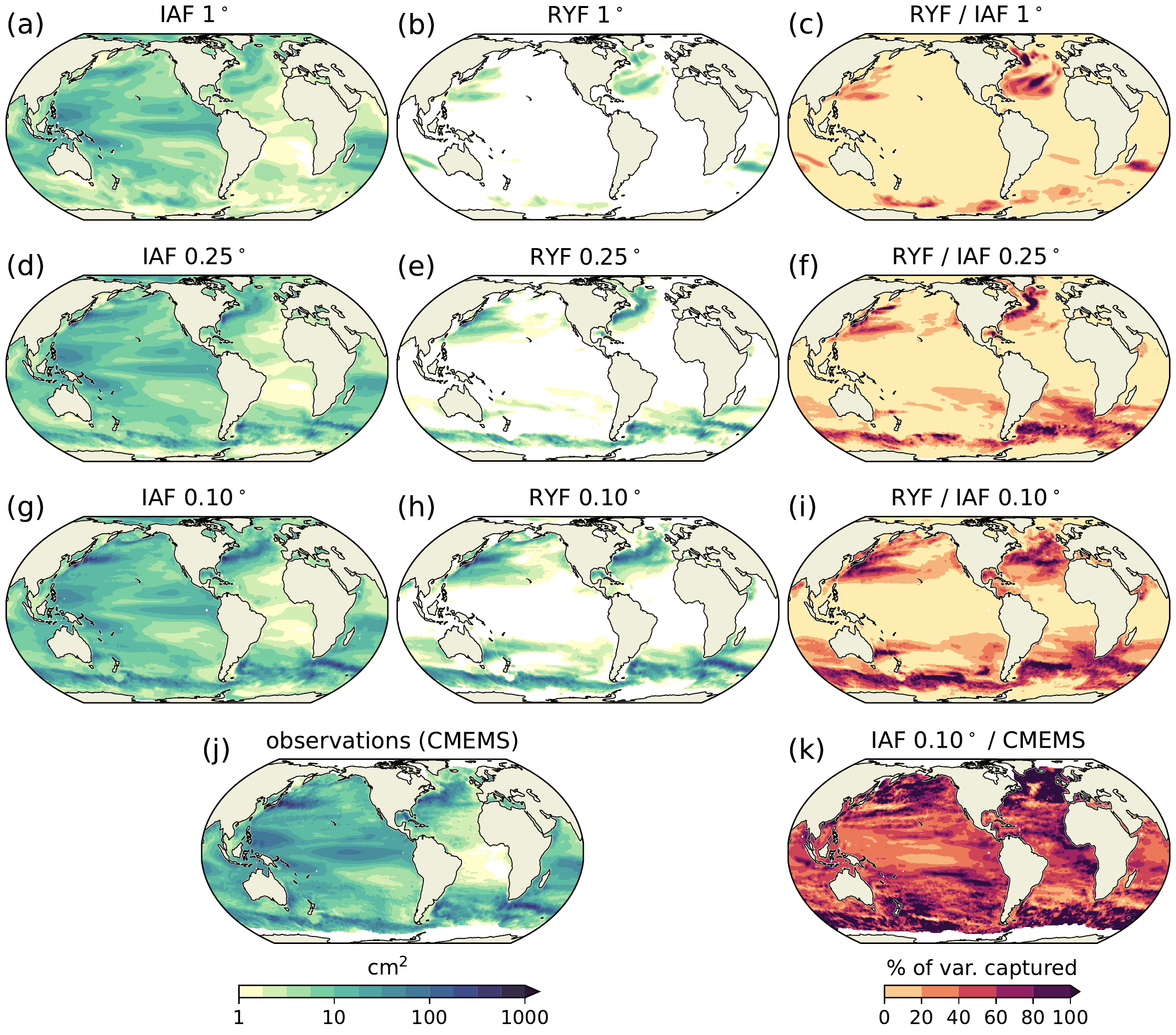}\\
 \caption{Sea-surface height (SSH) low-frequency variance, $\var{[\mathrm{SSH}_{\rm LF}]}$ (frequencies between $(25\,\textrm{years})^{-1}$ and $(1.5\,\textrm{years})^{-1}$) from ACCESS-OM2 models at three different resolutions and from satellite altimetry (CMEMS). 
 Panels~(a), (d), and~(g) show models forced with inter-annually varying forcing (IAF) from the JRA55-do dataset during 1958–2018; panels (b), (e), and~(f) show models forced with repeat-year forcing (RYF) from the JRA55-do dataset; panel (j) shows results from observations. 
 Panels (c), (f), and~(i) show the ratio of the RYF over the IAF low-frequency variance at the corresponding model resolution. 
 Panel~(k) shows the ratio of the low-frequency variance in observations over that from the IAF model at 0.10$^\circ$ for the same period. 
 (Only data between 65$^\circ$S and 65$^\circ$N were used from the CMEMS dataset.)}\label{fig1}
\end{figure*}

Figure~\ref{fig1} shows maps of the low-frequency variance for sea-surface height (see eq.~\eqref{eq:varHLF}). 
Panels~(a), (d), and~(g) show the results from the inter-annually forced models at the three different resolutions while panel~(j) shows the same map obtained from satellite altimetry (CMEMS). 
The low-frequency variance in the inter-annually forced model at high resolution (panel~g) resembles that in the observations (panel~j); panel (k) shows the ratio of the IAF 0.10$^\circ$ variance over observations (for the period 1993-2018), demonstrating that in most of extra-tropical regions the model captures at least 80\% of the low-frequency variance seen in observations. 
Close inspection of the ratio in panel~(k) reveals that the regions where low-frequency variance differs between model and observations are in the tropical Pacific and Indian Ocean, and the extension of the Kuroshio Current East of Japan. 
The tropical Indo--Pacific is dominated by interannual variability (El Ni\~no Southern Oscillation [ENSO] and Indian Ocean Dipole [IOD]), which depends on coupled ocean--atmosphere dynamics \citep{Webster-etal-1999,Timmermann-etal-2018}; therefore, weaker variability in this region is expected in an ocean-only simulation. 
Regarding the Kuroshio extension region, it has been documented that the Kuroshio Current undergoes decadal transitions between two different states \citep{Qiu-Chen-2010} and the variance associated with these transitions requires a longer observational record to be accurately captured. 

Comparison of the IAF simulations at the three different resolutions (panels~(a), (d), (g)) shows that, on one hand, the low-frequency variance in the tropics seems insensitive to increasing resolution, while on the other hand, mid-latitude and eddy-rich regions (like the Southern Ocean, and the Kuroshio Current, the Gulf Stream, and their extensions) show progressively more low-frequency variance as we increase the model resolution. 
Remember that flow fields from models across all resolutions have been interpolated to a grid with nominal 1$^\circ$ spacing and, therefore, the increased extra-tropical variance of panel~(g) compared with panel~(a) suggests that the model resolution affects the large-scale, low-frequency patterns of variability for the eddy-rich model, when compared with the eddy-present model. 
The insensitivity of the low-frequency variance in the tropics to model resolution may occur because variability in that region is dominated by atmospheric interaction, or because eddies, tropical instability waves and large-scale Kelvin waves, which come into play in the tropics and in particular in the tropical Pacific, are partially resolved even at the coarse 1$^\circ$ resolution.

\begin{figure*}[t]
 \noindent\centering\includegraphics[width=\textwidth, angle=0]{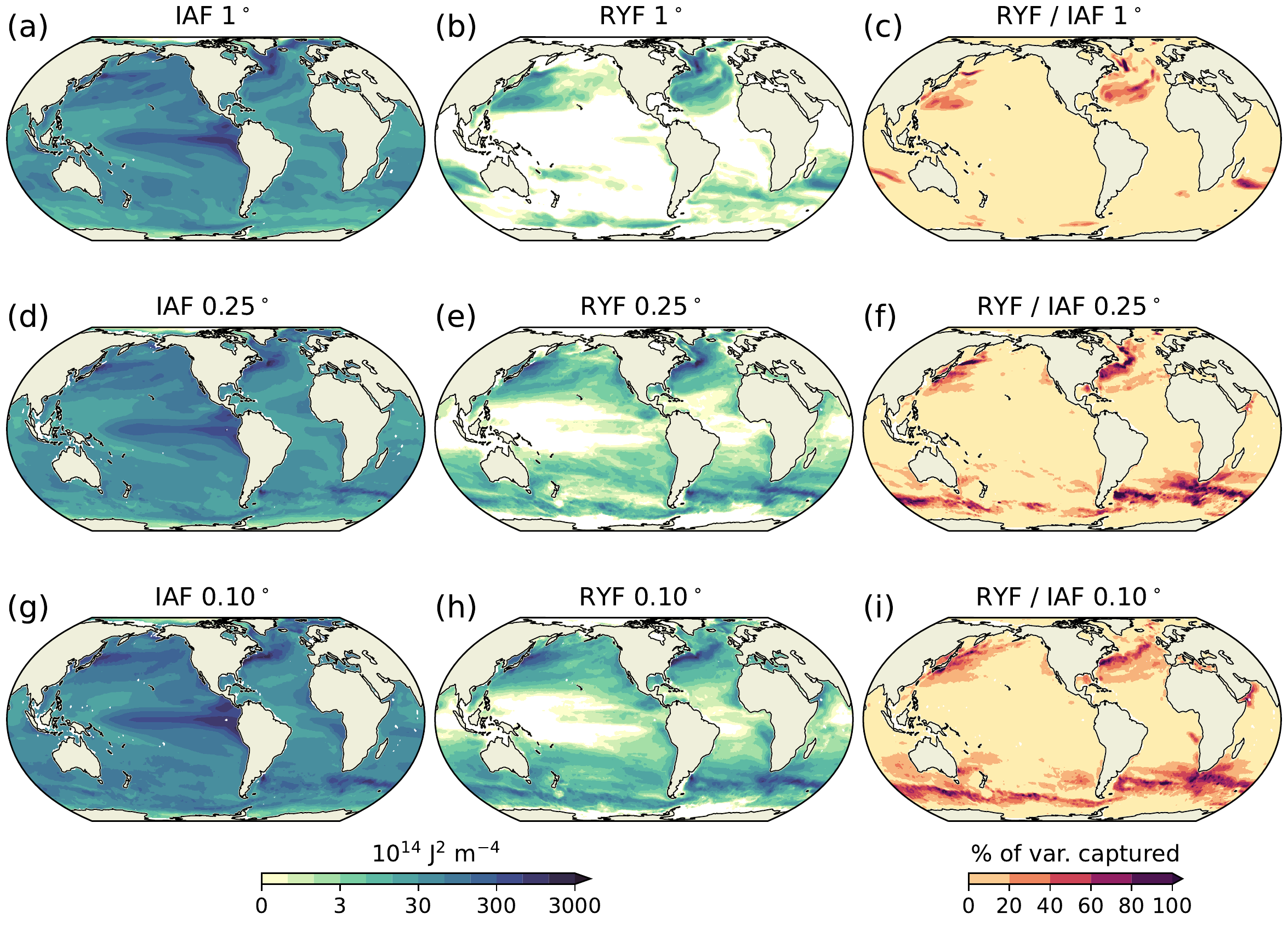}\\
 \caption{Upper-ocean heat content (top 50 meters) low-frequency variance, $\var{[\mathcal{H}_{\rm LF}]}$ (frequencies between $(25\,\textrm{years})^{-1}$ and $(1.5\,\textrm{years})^{-1}$)  from ACCESS-OM2 models at three different resolutions. 
 Left column (panels (a), (d), and (g)), shows models forced with inter-annually varying forcing (IAF) from JRA55-do reanalysis 1958-2018; middle column (panels (b), (e), and (h)) shows models forced with repeat-year forcing (RYF) from the JRA55-do reanalysis; right column (panels (c), (f), and (i)) shows the ratio of the RYF over the IAF frequency content at the corresponding model resolution.}\label{fig2}
\end{figure*}

Oceanic eddies are  generated intrinsically from oceanic processes (baroclinic or barotropic instability of currents, bathymetry interactions) and these processes are particularly active in the mid-latitudes. 
Thus, one would logically hypothesize that oceanic eddies are likely to contribute to increased variance at these latitudes. 
However, eddies occur on time scales less than a year, so the observed increase in \emph{low-frequency} variance between the 1$^\circ$ IAF model (panel (a)) and the 0.10$^\circ$ IAF model (panel (g)) cannot be explained by the direct effect of ocean eddies.

To better understand the contribution of the intrinsic oceanic response to the atmospheric forcing, and how that leads to decadal patterns of variability, we turn next to the repeat-year forcing model runs. 
The corresponding maps of the integrated low-frequency variance for sea-surface height for RYF models are shown in panels~(b), (e), and~(h) of figure~\ref{fig1}. 
The striking difference compared with the IAF model runs is the lack of low-frequency variance at the tropics. 
This lack of tropical low-frequency variability in the RYF runs implies that low-frequency variability in the tropics is dominated by atmospheric interactions (e.g., ENSO), and accounts for the insensitivity of tropical variability to resolution in the IAF simulations. 
On the other hand, the RYF models show significant low-frequency variance in the extra-tropical regions that increases with model resolution (compare panels (b), (e), and (h)). 
This variance is attributed to intrinsic oceanic variability. 

One can ask how much of the total low-frequency variance is due to intrinsic processes and how much is due to the low-frequency component of the forcing. 
Assuming that the IAF runs represent low-frequency variance resulting from both intrinsic and forced motions, then the ratio of the low-frequency content variance from RYF over that from IAF (panels (c), (f), and (i) respectively) depicts the percentage of intrinsic low-frequency variance in each model resolution. 
Two things should be taken away from panels (c), (f), and (i): first, that the percentage of intrinsic variance increases with model resolution; and second, that the regions of intrinsic low-frequency variance in models with parameterized eddies differ from those seen in the eddy-rich models.

The analysis of the sea-surface height above demonstrates that \emph{(i)}~the IAF models, at least at the eddy-rich 0.10$^\circ$ resolution, adequately capture the variance seen in the observational record from satellite altimetry (figure~\ref{fig1}(k)), and \emph{(ii)}~that intrinsic low-frequency variance is enhanced when oceanic eddies are resolved. 
Point~\emph{(i)} allows us to trust that our model captures the phenomena we are trying to assess, while point~\emph{(ii)} suggests that this analysis is worthy of further investigation. 
We turn now to the upper-ocean heat content, a quantity that may directly influence the atmosphere.

Figure~\ref{fig2} shows maps of the integrated low-frequency variance for upper-ocean heat content (in a similar manner as seen in figure~\ref{fig1} for sea-surface height, but without any comparison to observations). 
The results are similar to those seen in the sea-surface height analysis of figure~\ref{fig1}: intrinsic low-frequency variance increases with model resolution outside the tropics, while low-frequency variance within the tropics only appears in IAF model runs and is insensitive to model resolution. 
Furthermore, for the eddy-rich 0.10$^\circ$ resolution, panel~(i) suggests that in the eddy-rich regions, more than 50\% of the low-frequency variance is due to intrinsic oceanic variability.

\begin{figure*}[t]
 \noindent\centering\includegraphics[width=\textwidth, angle=0]{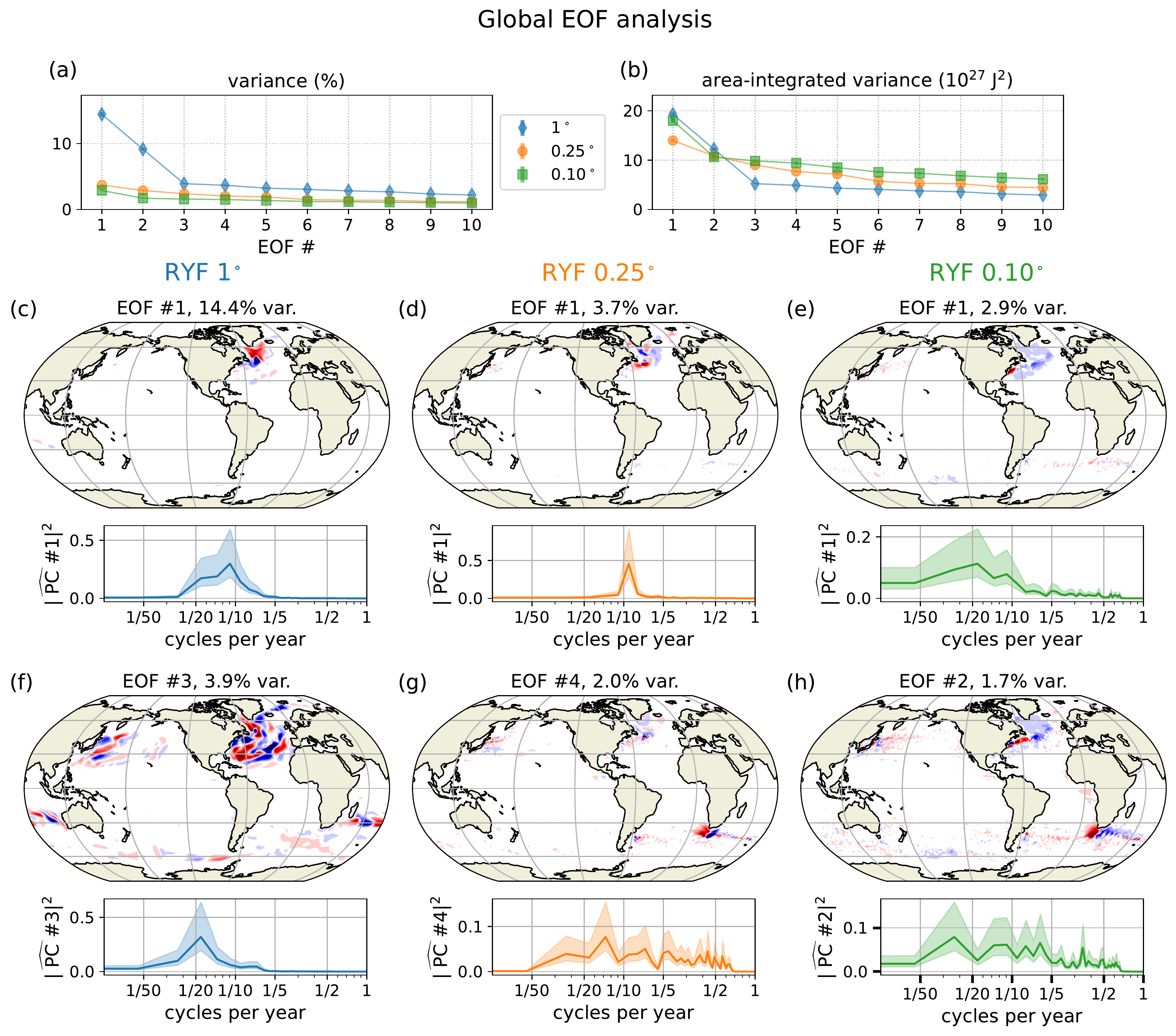}\\
 \caption{Global EOF analysis of the low-frequency reconstructed upper-ocean heat content, $\mathcal{H}_{\rm LF}$ (see eq.~\eqref{eq:H_LF}), for models driven by repeat-year forcing at 1$^\circ$, 0.25$^\circ$, and 0.10$^\circ$ resolutions. 
 Panels (a) and (b) show the percentage and the area-integrated explained variance for the first ten EOFs. 
 Panels (c)-(e) depict the first EOF for each model resolution, while panels (f)-(h) show another EOF, selected to emphasize a mode that is qualitatively different from EOF \#1. 
 Shown is the spatial structure of the EOF and the frequency power spectrum of the principal component (PC) time-series of the corresponding EOF. 
 Contours for the EOF maps in panels (c)-(h) are saturated at values $\pm\tfrac1{2}\max{|\text{EOF}|}$.}\label{fig3}
\end{figure*}

In both sea-surface height and upper-ocean heat content analyses, we note enhanced low-frequency variance in the North Atlantic, and in particular in the Labrador Sea, for the coarse 1$^\circ$-resolution model runs. 
This enhanced variance can be seen both in the IAF and RYF models (see figures~\ref{fig1}(a)-(b) and~\ref{fig2}(a)-(b)) but it is not apparent in eddy-rich model runs nor in the sea-surface height observations of figure~\ref{fig1}(j). 
This variability occurs because of spurious convection that occurs in the Labrador Sea in models with parameterized eddies \citep{Ortega-etal-2017}. 
The enhanced low-frequency variance that shows up in the Labrador Sea in 1$^\circ$ models is replaced by variance in the North Atlantic region at the location of the Gulf Stream extension in the 0.10$^\circ$ model.

\subsection{Spatial and temporal patterns of large-scale low-frequency variability}
\label{sec:eofs}

\begin{figure*}[t]
 \noindent\centering\includegraphics[width=\textwidth, angle=0]{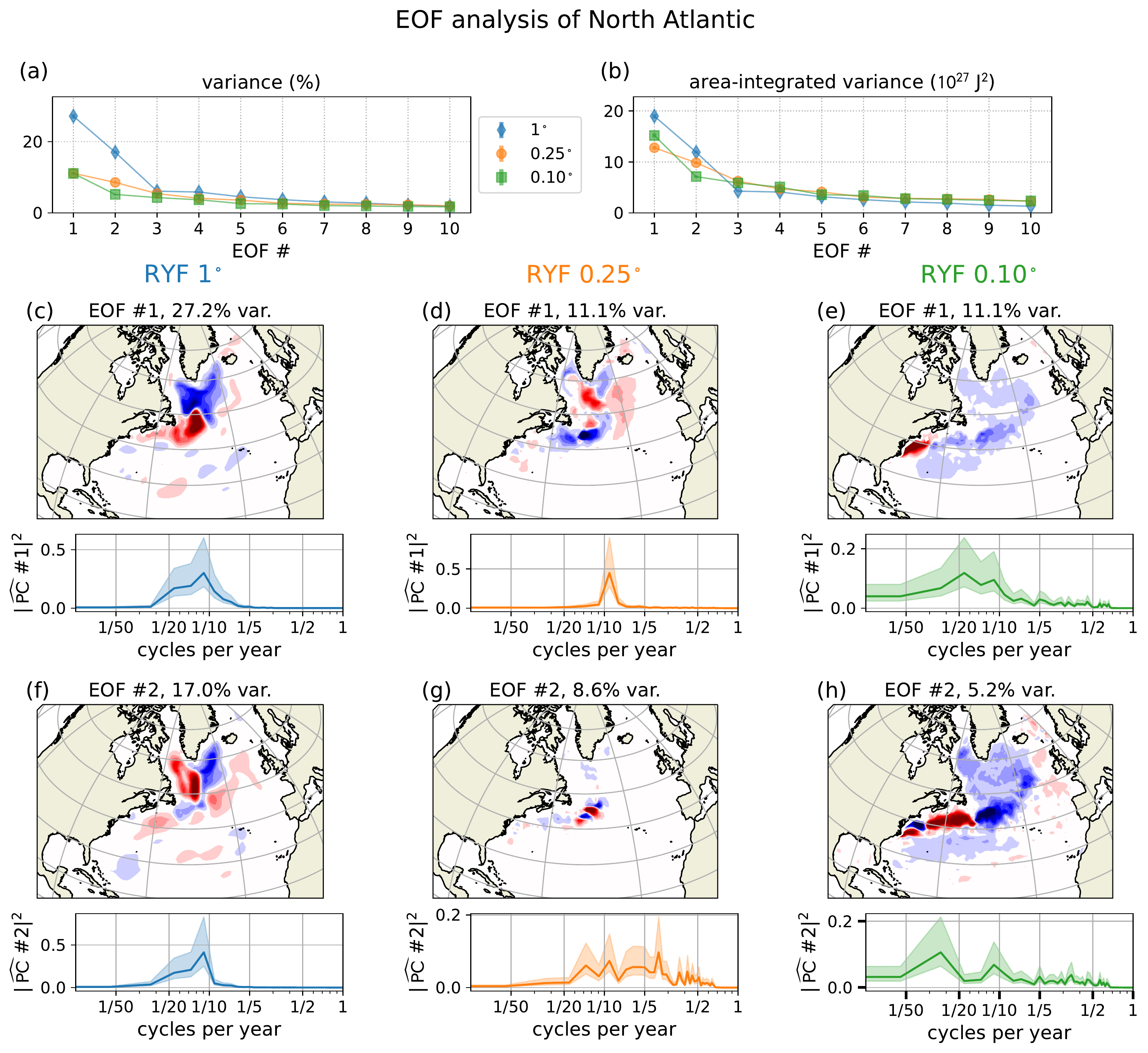}\\
 \caption{Regional EOF analysis for the North Atlantic Ocean (100$^\circ$W-0$^\circ$, 20$^\circ$N-75$^\circ$N). 
 (a)-(b)~Percentage and the area-integrated explained variance for the first ten EOFs over the region. 
 (c)-(h)~The spatial structure of the EOF and the frequency power spectrum of the principal component (PC) time-series of the corresponding EOF; shown are the first EOF for each model resolution and also an additional EOF selected to emphasize a mode that is qualitatively different from EOF~\#1. 
 Contours for the EOF maps in panels (c)-(h) are saturated at values $\pm\tfrac1{2}\max{|\text{EOF}|}$.}\label{fig4}
\end{figure*}

\begin{figure*}[t]
 \noindent\centering\includegraphics[width=\textwidth, angle=0]{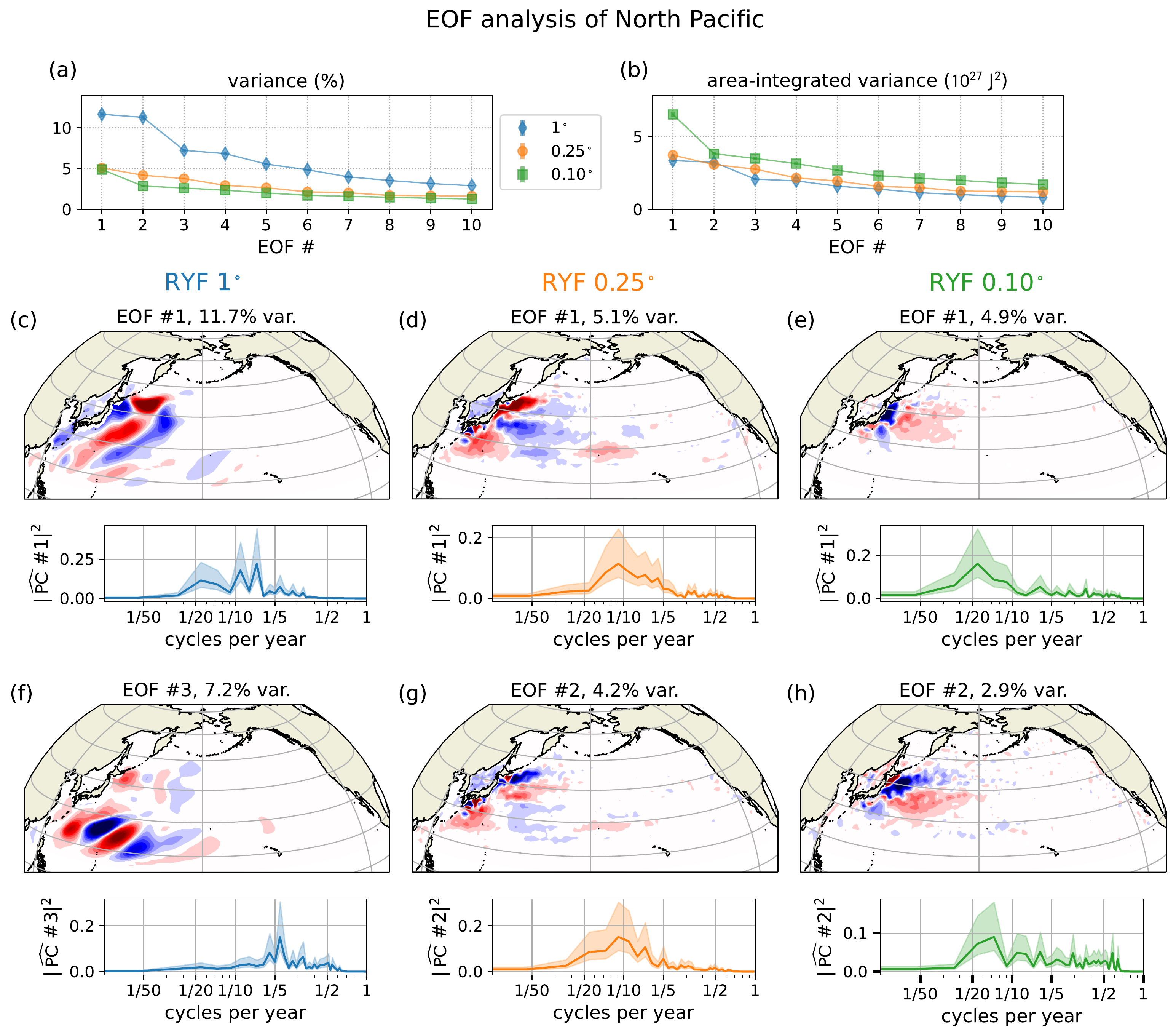}\\
 \caption{Regional EOF analysis for the North Pacific Ocean (90$^\circ$E-100$^\circ$W, 15$^\circ$N-70$^\circ$N). 
 (a)-(b)~Percentage and the area-integrated explained variance for the first ten EOFs over the region. 
 (c)-(h)~The spatial structure of the EOF and the frequency power spectrum of the principal component (PC) time-series of the corresponding EOF; shown are the first EOF for each model resolution and also an additional EOF selected to emphasize a mode that is qualitatively different from EOF~\#1. 
 Contours for the EOF maps in panels (c)-(h) are saturated at values $\pm\tfrac1{2}\max{|\text{EOF}|}$.}\label{fig5}
\end{figure*}

\begin{figure*}[t]
 \noindent\centering\includegraphics[width=\textwidth, angle=0]{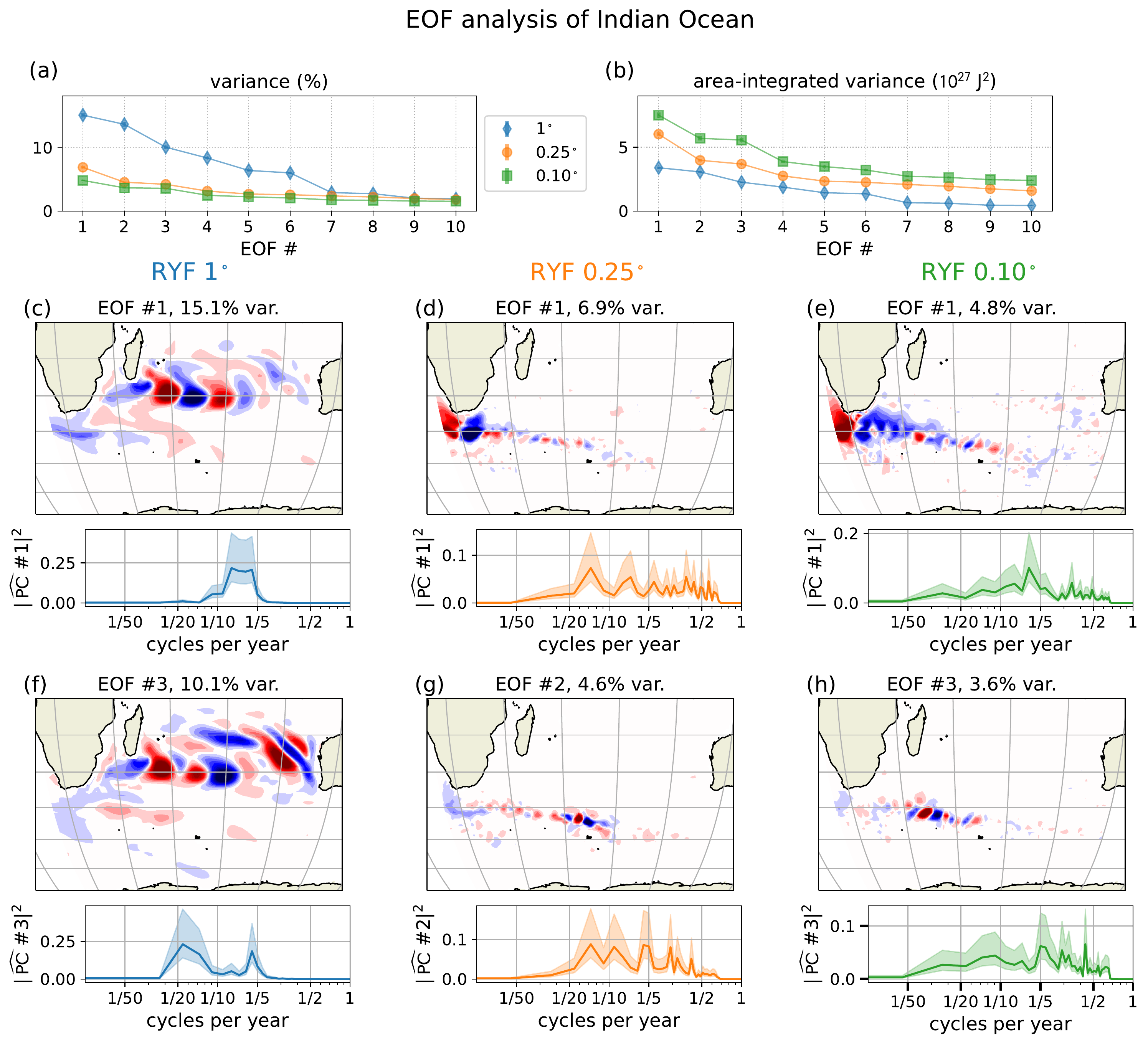}\\
 \caption{Regional EOF analysis for the Indian Ocean (5$^\circ$E-120$^\circ$E, 10$^\circ$S-70$^\circ$S). 
 (a)-(b)~Percentage and the area-integrated explained variance for the first ten EOFs over the region. 
 (c)-(h)~The spatial structure of the EOF and the frequency power spectrum of the principal component (PC) time-series of the corresponding EOF; shown are the first EOF for each model resolution and also an additional EOF selected to emphasize a mode that is qualitatively different from EOF~\#1. 
 Contours for the EOF maps in panels (c)-(h) are saturated at values $\pm\tfrac1{2}\max{|\text{EOF}|}$.}\label{fig6}
\end{figure*}

\begin{figure*}[t]
 \noindent\centering\includegraphics[width=\textwidth, angle=0]{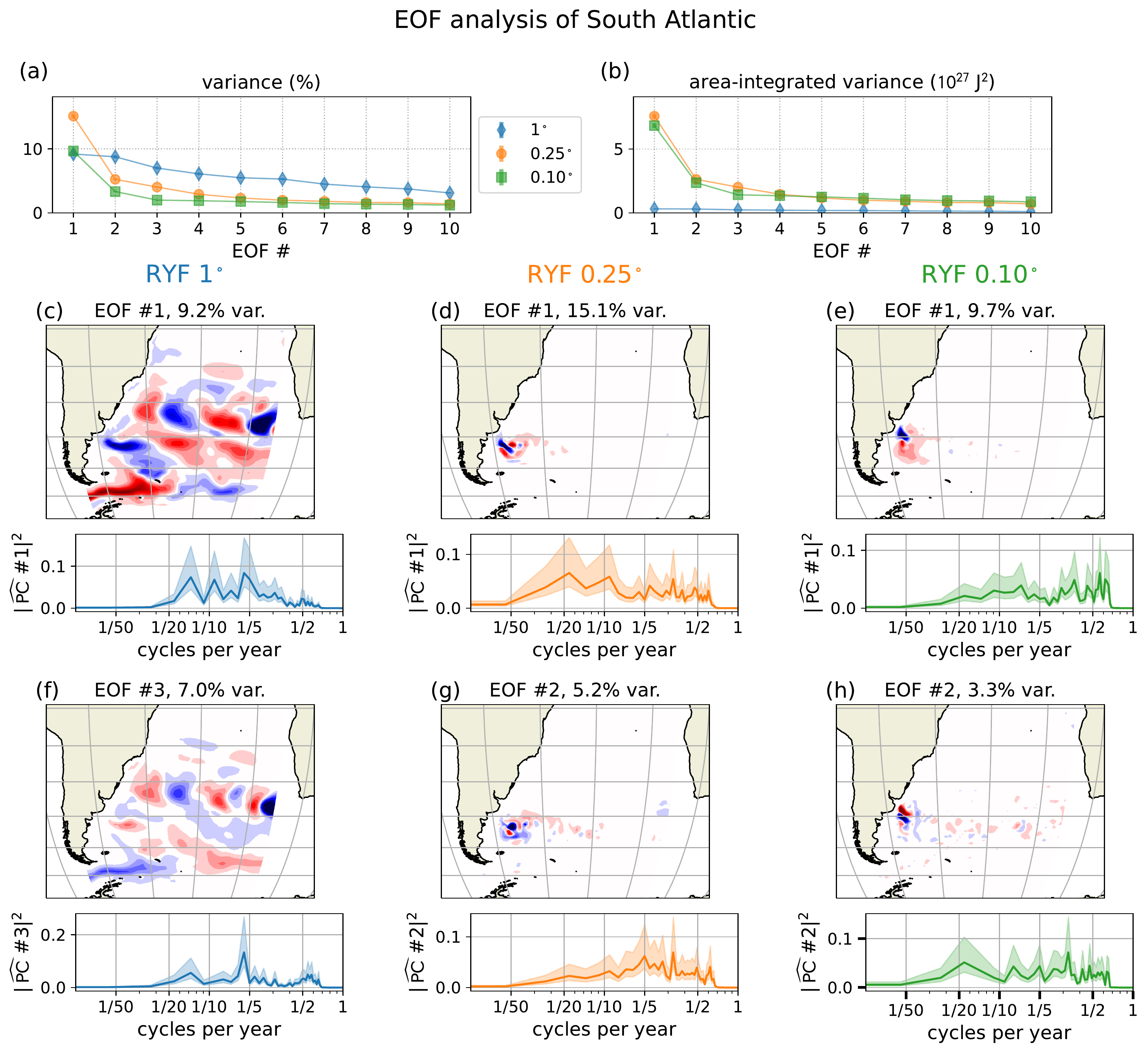}\\
 \caption{Regional EOF analysis for the South Atlantic Ocean (70$^\circ$W-10$^\circ$E, 10$^\circ$S-75$^\circ$S). 
 (a)-(b)~Percentage and the area-integrated explained variance for the first ten EOFs over the region. 
 (c)-(h)~The spatial structure of the EOF and the frequency power spectrum of the principal component (PC) time-series of the corresponding EOF; shown are the first EOF for each model resolution and also an additional EOF selected to emphasize a mode that is qualitatively different from EOF~\#1. 
 Contours for the EOF maps in panels (c)-(h) are saturated at values $\pm\tfrac1{2}\max{|\text{EOF}|}$.}\label{fig7}
\end{figure*}

We have established that  intrinsic oceanic variability leads to enhanced low-frequency patterns of upper-ocean heat content. 
But maps like those in figures~\ref{fig1} and~\ref{fig2}, only quantify the variance and do not elaborate on the spatial or temporal patterns of the low-frequency signal that the atmosphere may feel as a result of ocean's intrinsic variability. 
To better understand and quantify the spatial and temporal patterns of low-frequency intrinsic oceanic variability, we perform an Empirical Orthogonal Function (EOF)-analysis on the low-frequency reconstruction of the upper-ocean heat content, $\mathcal{H}_{\rm LF}$ (see eq.~\eqref{eq:H_LF}) from the RYF model output. 
Details on error bars and sensitivity checks of the EOF-analysis are provided in the \ifdraft Appendix\else\hyperref[appendix]{Appendix}\fi.

Figure~\ref{fig3} shows the global EOF analysis of the low-frequency upper-ocean heat content. 
Panel~(a) shows the percentage of the total low-frequency variance explained for each EOF and for each model resolution, while panel~(b) shows the total variance explained for each EOF for each model resolution.
Note that output from different model resolutions has been interpolated onto a grid of 1$^\circ$ nominal resolution. 
Therefore, the higher variance at increased resolution is not a direct effect of oceanic eddies. 
This result reiterates once more that models that resolve ocean eddies present enhanced large-scale, low-frequency variance of the upper-ocean heat content. 
The spatial patterns that emerge  (figure~\ref{fig3}(c--h)) are associated with the regions of enhanced variance seen in figure~\ref{fig2}(b), (e), and~(h). 
Note that the EOF patterns from the 0.1$^\circ$ model (figure~\ref{fig3}(e), (h)) resemble features associated with regions of high eddy activity.

The global EOF analyses in figure~\ref{fig3} are dominated by patterns of limited spatial extent. 
It follows that independent processes acting in different ocean basins meaning that we can refine our understanding of these individual processes by limiting the regional extent of these EOF analyses. 
Thus, we next perform regional EOF analyses in the North Atlantic, the North Pacific, the Indian Ocean, and the South Atlantic regions; see figures~\ref{fig4}-\ref{fig7} respectively. 
A feature that emerges from all EOF patterns is that, with increasing model resolution the patterns of low-frequency variability become more aligned with the eddy-rich regions, e.g., the western boundary currents and their extensions and the Southern Ocean. 
In terms of low-frequency sea-surface height variance, the highest resolution models showed better agreement with observations compared to coarse resolution models (see figure~\ref{fig1}(a), (d), (g), and~(j)). 
This result implies that the patterns concentrated in eddy-rich regions that were revealed by the EOF analysis for 0.10$^\circ$ resolution models are closer to reality compared with 1$^\circ$ resolution models.

In the North Atlantic, the 1$^\circ$ model shows more low-frequency variance than the higher resolution models. 
The variance the 1$^\circ$ model occurs in the Labrador Sea (figure~\ref{fig4}(c,f)) and it is intimately related to the spurious numerical convection that occurs there. 
We note that the signature of Labrador Sea spurious convection (figure~\ref{fig4}(c,f)) is almost completely eliminated in the eddy-rich 0.10$^\circ$ model (figure~\ref{fig4}(e,h)) which shows low-frequency variance at the location of the Gulf Stream and its extension. 
The leading EOF of the eddy-present model (RYF 0.25$^\circ$) appears as a combination of the other two, with patterns both in the Gulf Stream region and in the higher latitudes of the Labrador Sea, where the 0.25$^\circ$ is not adequate to resolve eddies.\footnote{The first baroclinic Rossby radius of deformation decreases as we move polewards, and, consequently, this results in smaller typical sizes for eddies closer to the poles \citep{Hallberg-2013}.} It is also interesting to note that the frequency power spectrum of the dominant EOFs show a shift towards lower frequencies as we increase model resolution. 
Therefore, this analysis demonstrates that the dominant effect of improving model resolution in the North Atlantic is to focus low-frequency variability in eddying regions and to lengthen the time scale of that variability.

In the North Pacific, substantial differences in the EOF patterns between the 1$^\circ$ model and the higher resolution models are found once again. 
The EOFs of the 1$^\circ$ model emerge as planetary waves traveling across the Pacific, while the higher resolution models show patterns associated with the Kuroshio Current. 
The eddy-rich 0.1$^\circ$ model shows patterns of variability with non-zonal character compared with the eddy-present 0.25$^\circ$ model. 
Again, the frequency power-spectrum of the principal components shows a shift towards lower frequencies for the eddy-rich, compared with the eddy-present, model. 
Furthermore, the spectral peak in figures~\ref{fig5}(e) and~(h) is between 15 and 25 years.

The patterns of variability in the Indian Ocean for the 1$^\circ$ model with parameterized eddies are planetary waves, while the patterns for the higher resolution models seem to be associated with the Agulhas retroflection and the Southern Ocean fronts (figure~\ref{fig6}). 
Long baroclinic Rossby wave patterns, similar to those seen in the subtropical Indian Ocean for the 1$^\circ$ model (figure~\ref{fig6}(c), (f)) and at the same location, dominate the variability in a 1$^\circ$-resolution coupled climate model \citep{Wolfe-etal-2017, Chapman-etal-2020}. 
That we find these patterns emerging out of the EOF analysis of the repeat-year forced ocean--sea ice model, argues that these long baroclinic Rossby waves are, at least to some extent, a result of ocean's intrinsic variability rather than from the coupled ocean--atmosphere interaction. 
We were not able, however, to see similar patterns emerging in the higher-resolution repeat-year forced models (RYF 0.25$^\circ$ and 0.10$^\circ$) in the first 30 EOFs. 
This result suggests that, at least the intrinsic oceanic component of the long baroclinic waves discussed by \cite{Wolfe-etal-2017} and \citet{Chapman-etal-2020}, is a feature of models with resolution that require oceanic eddies to be parameterized. 

Another feature of figure~\ref{fig6} is the similarity between the EOF patterns at eddy-present and eddy-rich resolutions. 
In both cases, the primary mode of variability is a broadband mode which manifests in variations in the zonal position of the Agulhas retroflection. 
The other mode shown in figure~\ref{fig6}(g),~(h) is lower frequency variability in the meander downstream of the Agulhas region. 
The similarity of these patterns suggests that 0.25$^\circ$ resolution may be sufficient to represent the primary modes of low-frequency variability in this region.

Lastly, the EOF analysis in the South Atlantic Ocean region (figure~\ref{fig7}) argues that the main pattern of variability is related to the Malvinas Current region, with the higher resolution models showing a tendency for a peak of the principal component power spectrum at longer time scales. 
For the 1$^\circ$ model, the low-frequency variance is almost negligible (figure~\ref{fig7}(b)) and, once again, planetary-wave patterns dominate the variability.

The North Pacific, South Atlantic, and Indian Ocean EOF analyses, reveal striking differences between the 1$^\circ$ model with parameterized ocean eddies and the other two resolutions. 
In these regions, the explained variance of the 1$^\circ$ is less than the eddy-present/eddy-rich models. 
The 1$^\circ$~EOFs show wave-like patterns, which may be a signature of baroclinic Rossby waves (see, e.g., figure~\ref{fig5}(c)). 
On the other hand, when eddies are at least partially resolved, these long waves are replaced with patterns that mirror the eddy effect on large-scale circulation patterns, e.g., the Kuroshio Current in figure~\ref{fig5}(e).

\section{Discussion}
\label{sec:discussion}

The ocean's larger heat capacity, compared with that of the atmosphere, is commonly viewed as the cause of the ocean acting as an ``integrator'' for atmospheric high-frequency synoptic variability, reddening the power-spectrum of motions, i.e., enhancing power at low frequencies, that is with decadal time scales (null-hypothesis; \citet{Hasselmann-1976, Frankignoul-Hasselmann-1977}). 
In this work, we investigated whether the ocean's intrinsic dynamics, which give rise to the rich high-frequency mesoscale and submesoscale oceanic eddy features, can further enhance the low-frequency variability of the ocean. 
In particular, we have focused on whether higher-resolution ocean simulation show differences in the low-frequency variability of the upper-ocean heat content at length scales comparable to, or larger than the typical scales of atmospheric variability. 
Such low-frequency patterns of upper-ocean heat content may directly feed back on the atmosphere, give rise to decadal modes of climate variability and, therefore, potentially affect global climate.

We have assessed the effect of the intrinsic oceanic variability using a global ocean--sea ice model at three different resolutions: eddy-rich $0.10^\circ$, eddy-present 0.25$^\circ$ and with parameterized eddies at 1$^\circ$. 
To disentangle the forced from the intrinsic component of ocean's variability, we used two different forcing schemes: \emph{(i)} inter-annually varying forcing (IAF) from the JRA55-do atmospheric reanalysis 1958-2018 and \emph{(ii)} repeat-year forcing (RYF) from May 1990--April 1991 from the JRA55-do. 
Since the RYF does not force any time scales longer than 1 year, it allows us to directly probe the ocean's intrinsic variability at decadal time scales. 
We have used two fields to investigate large-scale, low-frequency patterns of intrinsic ocean variability: the upper-ocean heat content (top 50 meters of the ocean) and the sea-surface height. 
The sea-surface height served as a proxy for upper-ocean heat content and further allowed us to ground-truth our model results against observations from the satellite altimetry record. 
Despite the short length of the altimetry record (27 years), which does not capture well the decadal variability, the comparison of the low-frequency variance from the inter-annually forced eddy-rich model (0.10$^\circ$) model with that from the observations is encouraging (see figures~\ref{fig1}(g), (j), and~(k)).

Our results demonstrate that \emph{(i)}~models that resolve eddies have much more variance of the upper-ocean heat content at decadal time scales (section~\ref{sec:results}\ref{sec:varmaps}) and, furthermore,~\emph{(ii)} the spatial patterns of the low-frequency variability of upper-ocean heat content are limited in models that rely on eddy parameterizations (see section~\ref{sec:results}\ref{sec:eofs}). 
The direct effect of the intrinsic ocean dynamics on the low-frequency variability of upper-ocean heat content is negligible in the tropical regions but it is particularly pronounced in the extratropics, where the ocean's mesoscale is most active (see figure~\ref{fig1} and~\ref{fig2}). 
Eddy-resolving models not only show enhanced low-frequency variance but, furthermore, EOF-analysis reveals that the main modes of variability have patterns that are aligned with nonlinear flow structures, rather than baroclinic waves or spurious convection. 
Last, variability occurs at lower frequencies in higher resolution models. 

The differences in the low-frequency variability across model resolutions could be attributed either to changes on the mean states that the models show (e.g., mean currents or mean stratification), or due to dynamical processes that are enabled with higher resolution. 
The mean state of these models across 1$^\circ$, 0.25$^\circ$, and 0.10$^\circ$ resolution is not substantially different \citep{Kiss-etal-2020} (except, of course, in the Labrador Sea region where the 1$^\circ$ models show spurious convection). 
We, therefore, argue that the most likely candidates to explain the changes in the patterns and the changes in the timescales of the low-frequency variability we observe are eddy--mean flow interactions (as, for example, the case of the turbulent oscillator; \citep{Berloff-etal-2007}), wave--eddy interactions \citep{Serazin-etal-2018}, or eddy--eddy interactions. 

The above-mentioned results may be used to infer a role for the extra-tropical ocean in generating modes of low-frequency climate variability such as the North Atlantic Oscillation \citep[NAO; ][]{hurrell-etal-2001} or the Interdecadal Pacific Oscillation \citep[IPO; ][]{Mantua-Hare-2002}. 
For example, in the Pacific Ocean, the time scale of dominant modes of variability for the eddy-rich model at $0.10^\circ$ is between 10 to 20~years (see figure~\ref{fig5}(e), (h)), which is similar to the time scale of the IPO. 
Recent Earth system simulations revealed that the IPO is in better agreement with observations in simulations with higher lateral resolution \citep{Chang-etal-2020}. 
The IPO is a much broader pattern of variability than the EOF patterns we see in figure~\ref{fig5}. 
However, a viable hypothesis may be that intrinsic oceanic variability in the extratropics could be amplified at the surface by ocean--atmosphere feedback mechanisms such as the wind-evaporation-sea-surface-temperature feedback \citep{Xie-Philander-1994} and conveyed to the tropics through dynamics known as Pacific Meridional Modes \citep{Alexander-etal-2010, DiLorenzo-etal-2015, Amaya-2019}. 
Once they propagate into the tropics these sea-surface-temperature anomalies can modulate ENSO dynamics at interdecadal frequencies and feed back to the extratropics via ENSO teleconnections \citep{Newman-etal-2003,Alexander-etal-2002}, hence explaining the broader signature of the main low-frequency mode of variability of the Pacific Ocean, the IPO. 
In summary, the results we presented in this paper argue that \emph{(i)}~that oceanic dynamics are largely responsible for the the better agreement compared to observations seen in higher resolution simulations and \emph{(ii)}~we further speculate that the extratropical oceans may provide the low-frequency trigger that might help to explain the IPO.

In view of the discussion above, we conclude that resolving the oceanic eddies in the eddy-active regions, and thus better capturing the low-frequency variability induced by intrinsic ocean dynamics in the extra-tropics, has global implications for modeling decadal variability and hence decadal predictions and climate projections. 
These results may have ramifications for our interpretation of model output and, in turn, affect future climate predictions. 
The large majority of CMIP6 climate models used for climate projections have an ocean component with a 1$^\circ$ resolution that is too coarse to resolve ocean eddies. 
Our results, thus, suggest that in those CMIP6 climate models, the ocean's input to atmospheric variability is weaker (compare figures~\ref{fig2}(b) and~(h)) and also has different spatial patterns than when eddies are represented (compare, e.g., figures~\ref{fig3}(c), (e) with figures~\ref{fig3}(e), (h)). 
Correctly capturing these large-scale, low-frequency patterns of variability that force the atmosphere is crucial for predicting climate modes of variability at decadal time scales (El Ni\~no, Interdecadal Pacific Oscillation, North Atlantic Oscillation, Indian Ocean Dipole) but, also, crucial for future climate predictions. 
Our results suggest the imperative for moving towards coupled climate models with ocean component with, at least, an eddy-present lateral resolution.








\acknowledgments

\begin{figure}[h!]
\centering
\noindent\includegraphics[width=16pc,angle=0]{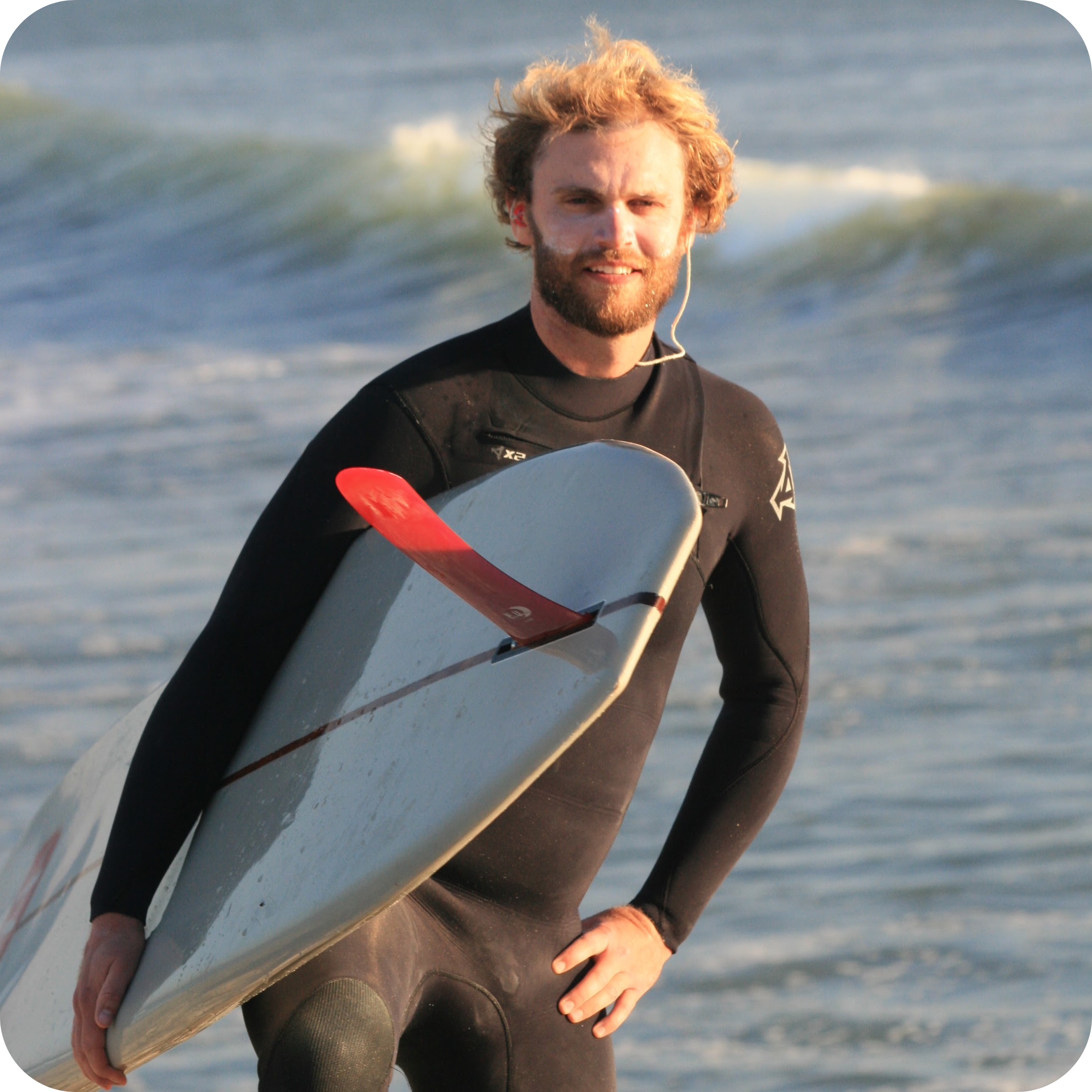}\\
\caption{Sean R. Haney; February 1987--January 2021.}\label{sean}
\end{figure}

Our analyses were facilitated with the Python packages \texttt{dask} \citep{Rocklin-2015}, \texttt{eofs} \citep{Dawson-2016}, \texttt{xarray} \citep{Hoyer-Hamman-2017}, \texttt{xESMF} \citep{xESMF-v0.5.2}, and \texttt{xrft} \citep{xrft-zenodo-v0.2.1}. 
Computational resources were provided by the Australian National Computational Infrastructure at ANU, which is supported by the Commonwealth of Australia. 
The authors would like to thank the three anonymous reviewers and the editor for their constructive comments that helped improving the manuscript. 
Discussions with Leela Frankcombe, Paige Martin, Thierry Penduff, and with members of the Consortium for Ocean--Sea Ice Modeling in Australia (\href{http://cosima.org.au}{www.cosima.org.au}) are greatly acknowledged.
N.C.C.~would like to further thank Chris Chapman for discussions regarding baroclinic planetary waves in the Indian Ocean, Rishav Goyal for discussion regarding EOF analysis, Petros Ioannou for discussions regarding bi-stability in geophysical flows, Giovanni Liguori for discussions regarding teleconnection patterns between the tropics and extra-tropics, Paige Martin and Takaya Uchida for their help with the python package \texttt{xrft}, and Andreas Selamtzis for discussions regarding signal processing. 
N.C.C. is grateful to his friend Andy~H. from Cook, Australian Capital Territory for the support he provided him during the solitary times of the Covid-19 lock-down in Canberra: the hours they spent together wood working were balsam to the soul.
N.C.C. is also indebted to the people of Broulee and its surroundings at the South Coast of New South Wales for encompassing him into their community from day one: embraced by ``Da Crew'' I have found  a new home away from home. 
Last, N.C.C. would like to take a moment to remember physical oceanographer Sean R.~Haney (figure~\ref{sean}) who sadly passed away while this paper was in review: Sean was my office mate at the Scripps Institution of Oceanography, my surfing instructor, and a best friend.

%
%
\datastatement
The satellite altimetry products were produced by Ssalto/Duacs and distributed by E.\;U.~Copernicus Marine Service Information at \href{https://resources.marine.copernicus.eu/?option=com_csw&task=results&pk_vid=58f1a8b4b3c3a02e1612408547b5747e}{https://resources.marine.copernicus.eu}. 
Some of the raw ACCESS-OM2 model output is available at doi:\href{https://doi.org/10.4225/41/5a2dc8543105a}{10.4225/41/5a2dc8543105a}. 
All the model output that was used for this paper is available at Zenodo repository, doi:\href{https://doi.org/10.5281/zenodo.4924968}{10.5281/zenodo.4924968}. 
All figures and analyses in this paper can be reproduced using the Jupyter notebooks that are publicly available at the GitHub repository \href{https://github.com/navidcy/IntrinsicOceanicLFVariabilityUOHC}{github.com/navidcy/IntrinsicOceanicLFVariabilityUOHC}.
%
\appendix


\appendixtitle{EOF Analysis: error bars and robustness} \label{appendix}

The standard error of the EOF-analysis eigenvalues is computed using the criterion by \citet{North-etal-1982}. 
This standard error is shown in both panels~(a) and~(b) in figures~\ref{fig3}--\ref{fig7}. 
It is generally small enough to be barely distinguishable over the marker size.

For the power spectra shown in panels~(c) to~(h) in figures~\ref{fig3}--\ref{fig7}, we split the principal-component time series into four consecutive segments (55 year-long at least) and then computed the mean power spectrum of of each segment. 
The error bars were obtained using the $\chi^2$ criterion with a 95\% confidence interval.

The robustness of the spatial patterns of the EOF analysis was checked by comparing the patterns of EOF analysis from \emph{(i)}~full times series, \emph{(ii)}~the first 60\% of the time series, and \emph{(iii)}~the last 60\% of the time series. 
For the regions presented here we, typically, find that the EOF patterns \#1 and \#2 were found to be the same. 
Sometimes EOF patterns \#3-\#5 come in different order (e.g., what was EOF \#3 using the full time-series could be EOF \#4 when using the first 60\% of the time-series).

Jupyter notebooks that reproduce the EOF-analysis sensitivity tests can be found at the GitHub repository \href{https://github.com/navidcy/IntrinsicOceanicLFVariabilityUOHC}{github.com/navidcy/IntrinsicOceanicLFVariabilityUOHC}.




%
%
%

%


\end{document}